\documentstyle[12pt]{article}

\makeatletter
\advance \topmargin by -\headheight
\advance \topmargin by -\headsep
\evensidemargin \oddsidemargin
\def\lesssim{\mathrel{\lower2.5pt\hbox{$\textstyle<$}\atop\raise2.5pt\hbox{$\textstyle\sim$}}}

\def\widetext{}

\def\pacs#1{}
\def\narrowtext{}
\def\tightenlines{}

\def\@authoraddress{}  \def\@title{} \def\@date{} \def\@preprint{}
\def\and{\unskip, }
\def\preprint#1{%
\def\@preprint{\noindent\hfill\hbox{#1}\vskip 10pt}%
}
\def\title#1{\gdef\@title{{\large\bf\centering\ignorespaces#1\vskip2.5pt}}}
\def\author#1{\expandafter\def\expandafter\@authoraddress\expandafter
{\@authoraddress %
\vskip1.5pc %
{\dimen0=-\prevdepth \advance\dimen0 by23pt
\nointerlineskip \rm\centering
\vrule height\dimen0 width0pt\relax\ignorespaces#1\par
}%
}%
}
\def\address#1{\expandafter\def\expandafter\@authoraddress\expandafter
{\@authoraddress{\small\it\centering\ignorespaces#1\par}}}
\def\date#1{\gdef\@date{{\small\rm\centering(\ignorespaces#1\unskip)\par}}}
\def\maketitle{\par
\begingroup
\let\cite\@bylinecite
\def\thefootnote{\fnsymbol{footnote}}%
\if@twocolumn
\twocolumn[\@maketitle\vskip2pc]%
\else
\newpage
\global\@topnum\z@ %
\@maketitle
\fi
\thispagestyle{plain}\@thanks
\endgroup
\def\thefootnote{\arabic{footnote}}%
\setcounter{footnote}{0}%
\let\maketitle\relax \let\@maketitle\relax
\let\@thanks\relax \let\@authoraddress\relax \let\@title\relax
\let\@date\relax \let\thanks\relax
}
\def\@maketitle{%
\@preprint
\@title
\ifdim\prevdepth=-1000pt \prevdepth0pt\fi
\@authoraddress
\@date
}
\def\thesection       {\Roman{section}}
\def\p@section        {}
\def\thesubsection    {\Alph{subsection}}
\def\p@subsection     {\thesection\,}

\def\p@subsubsection  {\thesection\,\thesubsection\,}

\def\acknowledgments{\section*{ACKNOWLEDGMENTS}}

\newif\if@mainhead
\let\reset@font\relax
\def\section{\@mainheadtrue
\@startsection {section}{1}{\z@}{0.8cm plus1ex minus
 .2ex}{0.5cm plus1ex minus.2ex}{\reset@font\small\bf\centering}}
\def\subsection{\@mainheadfalse
\@startsection{subsection}{2}{\z@}{0.8cm plus1ex minus
 .2ex}{0.5cm plus1ex minus.2ex}{\reset@font\small\bf\centering}}
\def\subsubsection{\@mainheadfalse
\@startsection{subsubsection}{3}{\z@}{.8cm plus1ex minus
 .2ex}{0.5cm plus1ex minus.2ex}{\reset@font\small\it\centering}}
\def\paragraph{\@mainheadfalse
\@startsection{paragraph}{4}{\parindent}{\z@}{-1em}{\reset@font
\normalsize\it}}
\def\subparagraph{\@mainheadfalse
\@startsection{subparagraph}{4}{\parindent}{3.25ex plus1ex minus
 .2ex}{-1em}{\reset@font\normalsize\bf}}

\def\baselinestretch{1.5}
\def\setstretch#1{\renewcommand{\baselinestretch}{#1}}
\@ifundefined{selectfont}
{
\def\@setsize#1#2#3#4{\@nomath#1
   \let\@currsize#1\baselineskip
   #2\baselineskip\baselinestretch\baselineskip
   \parskip\baselinestretch\parskip
   \setbox\strutbox\hbox{\vrule height.7\baselineskip
      depth.3\baselineskip width\z@}
   \normalbaselineskip\baselineskip#3#4}
}
{
\def\@setsize#1#2#3#4{\@nomath#1%
   \let\@currsize#1\parskip\baselinestretch\parskip
    \baselineskip\baselinestretch\baselineskip
              \size{#4}{#2}\selectfont}
}
\skip\footins 20pt plus4pt minus4pt

\def\@xfloat#1[#2]{\ifhmode \@bsphack\@floatpenalty -\@Mii\else
   \@floatpenalty-\@Miii\fi\def\@captype{#1}\ifinner
      \@parmoderr\@floatpenalty\z@
    \else\@next\@currbox\@freelist{\@tempcnta\csname ftype@#1\endcsname
       \multiply\@tempcnta\@xxxii\advance\@tempcnta\sixt@@n
       \@tfor \@tempa :=#2\do
                        {\if\@tempa h\advance\@tempcnta \@ne\fi
                         \if\@tempa t\advance\@tempcnta \tw@\fi
                         \if\@tempa b\advance\@tempcnta 4\relax\fi
                         \if\@tempa p\advance\@tempcnta 8\relax\fi
         }\global\count\@currbox\@tempcnta}\@fltovf\fi
    \global\setbox\@currbox\vbox\bgroup
    \def\baselinestretch{1}\small\normalsize
    \boxmaxdepth\z@
    \hsize\columnwidth \@parboxrestore}
\long\def\@footnotetext#1{\insert\footins{\def\baselinestretch{1}\footnotesize
    \interlinepenalty\interfootnotelinepenalty
    \splittopskip\footnotesep
    \splitmaxdepth \dp\strutbox \floatingpenalty \@MM
    \hsize\columnwidth \@parboxrestore
   \edef\@currentlabel{\csname p@footnote\endcsname\@thefnmark}\@makefntext
    {\rule{\z@}{\footnotesep}\ignorespaces
      #1\strut}}}
\def\singlespace{%
\vskip\parskip%
\vskip\baselineskip%
\def\baselinestretch{1}%
\@ifundefined{selectfont}{%
\ifx\@currsize\normalsize\@normalsize\else\@currsize\fi%
}{
\ifx\@currsize\normalsize\@normalsize\else\@currsize\fi\setnew@baselineskip%
}
\vskip-\parskip%
\vskip-\baselineskip%
}

\def\spacing#1{\par%
 \def\baselinestretch{#1}%
 \ifx\@currsize\normalsize\@normalsize\else\@currsize\fi}

\everydisplay{
   \abovedisplayskip \baselinestretch\abovedisplayskip%
   \belowdisplayskip \abovedisplayskip%
   \abovedisplayshortskip \baselinestretch\abovedisplayshortskip%
   \belowdisplayshortskip  \baselinestretch\belowdisplayshortskip}

\def\citen#1{%
\if@filesw \immediate \write \@auxout {\string \citation {#1}}\fi
\@tempcntb\m@ne \let\@h@ld\relax \def\@citea{}%
\@for \@citeb:=#1\do {%
  \@ifundefined {b@\@citeb}%
    {\@h@ld\@citea\@tempcntb\m@ne{\bf ?}%
    \@warning {Citation `\@citeb ' on page \thepage \space undefined}}%
    {\@tempcnta\@tempcntb \advance\@tempcnta\@ne
    \setbox\z@\hbox\bgroup 
    \ifnum0<0\csname b@\@citeb \endcsname \relax
       \egroup \@tempcntb\number\csname b@\@citeb \endcsname \relax
       \else \egroup \@tempcntb\m@ne \fi
    \ifnum\@tempcnta=\@tempcntb 
       \ifx\@h@ld\relax 
          \edef \@h@ld{\@citea\csname b@\@citeb\endcsname}%
       \else 
          \edef\@h@ld{\hbox{--}\penalty\@highpenalty
            \csname b@\@citeb\endcsname}%
       \fi
    \else   
       \@h@ld\@citea\csname b@\@citeb \endcsname
       \let\@h@ld\relax
    \fi}%
 \def\@citea{,\penalty\@highpenalty\hskip.13em plus.1em minus.1em}%
}\@h@ld}
\def\cite{\leavevmode\unskip\@ifnextchar[{\@tempswatrue\@citew}%
            {\@tempswafalse\@citex}}
\def\@citew[#1]#2{\ifnum\lastpenalty=\z@ \penalty\@highpenalty \fi
   \ [{\multiply\@highpenalty 3 
   \citen{#2}},\penalty\@highpenalty\ #1]\spacefactor\@m}
\def\@citex#1{\begingroup\leavevmode\unskip
  \def\@tempa{\@cite{\citen{#1}}\endgroup}\futurelet\@tempb\@citey}%
\def\@citey{\let\@tempc\@tempa
   \ifx\@tempb.\ifnum\spacefactor>2999 \let\@tempb\relax\fi\let\@tempc\@citez
   \else\ifx\@tempb,\let\@tempc\@citez
   \else\ifx\@tempb:\let\@tempc\@citez 
   \else\ifx\@tempb;\let\@tempc\@citez 
   \fi\fi\fi\fi
   \@tempc}%
\def\@citez#1{\@tempb\futurelet\@tempb\@citey}%
\def\@cite#1{$\m@th\the\scriptfont\z@\edef\bf{\the\scriptfont\bffam}%
      ^{\hbox{#1}}$}

\makeatother

\begin{document}

\preprint{cond-mat/9404001}

\title{Folding Kinetics of Protein Like Heteropolymers}

\author{Nicholas D. Socci and Jos\'e Nelson Onuchic}

\address{Department of Physics\\
         University of California at San Diego\\
         La Jolla, California 92093-0319\\
         {\rm Email:\quad{\tt nsocci@ucsd.edu}}\\[.25in]%
         \sf Accepted in J.\ Chem.\ Phys.}

\date{March 17, 1994}

\maketitle

\begin{abstract}

\tightenlines

  Using a simple three-dimensional lattice copolymer model and Monte
  Carlo dynamics, we study the collapse and folding of protein-like
  heteropolymers. The polymers are 27 monomers long and consist of two
  monomer types. Although these chains are too long for exhaustive
  enumeration of all conformations, it is possible to enumerate all
  the maximally compact conformations, which are
  {$3\!\times\!3\!\times\!3$} cubes.  This allows us to select
  sequences that have a unique global minimum. We then explore the
  kinetics of collapse and folding and examine what features determine
  the various rates. The folding time has a plateau over a broad range
  of temperatures and diverges at both high and low temperatures. The
  folding time depends on sequence and is related to the amount of
  energetic frustration in the native state. The collapse times of the
  chains are sequence independent and are a few orders of magnitude
  faster than the folding times, indicating a two-phase folding
  process. Below a certain temperature the chains exhibit glass-like
  behavior, characterized by a slowing down of time scales and loss of
  self-averaging behavior. We explicitly define the glass transition
  temperature ($T_g$), and by comparing it to the folding temperature
  ($T_f$), we find two classes of sequences: good folders with
  $T_f>T_g$ and non-folders with $T_f<T_g$.

\end{abstract}

\newpage\narrowtext

\section{INTRODUCTION}

It has been known for some time that for many proteins the information
necessary to specify the native structure is contained within the
amino acid sequence. There has been a tremendous amount of research
aimed at deciphering this code and determining the final structure
from the sequence. Solving this problem is of paramount importance;
however, simply knowing how to map sequences to structures would leave
many interesting questions unanswered. How do proteins fold to their
native structure and, more specifically, how do they manage to fold so
quickly? What are the key factors that determine whether or not a
given sequence will fold and what the folding time will be? One may
argue that it might be necessary to solve these problems before it
will be possible to solve the folding problem (i.e., predicting
structure from sequence).

A great deal of work (both experimental and theoretical) has been done
on the kinetics of protein folding. One extremely useful theoretical
technique is to study simple heteropolymer models. The idea is to
reduce the complex system of proteins in solution to its bare
essentials, leaving only the key features. The advantage of studying
these simpler models is that an in-depth analysis (sometimes even an
exhaustive one) can be performed, yielding detailed answers and
information. This information should, in turn, provide insights into
real proteins.

One class of model that is often used in theoretical polymer work is
the lattice model, where the monomers are constrained to lie on
lattice sites. Excluded volume is included by allowing only one
monomer per site. To study dynamics, the Monte Carlo algorithm with a
variety of move sets is used. Some of the earliest work using lattice
models on proteins was done by G\=o and others\cite{Ueda78,Go83a}
using two- and three-dimensional lattices to examine the folding
process. However, the interaction potential they used was somewhat
unusual. The native state was explicitly built into the potential. The
energy of any given conformation was determined by counting the number
of native contacts, {\em i.e.}, contacts found in the native
structure. An attractive contribution to the energy was added for each
native contact formed. This potential is somewhat unphysical,
depending on an {\em a priori} knowledge of the native structure.
Although much of this early work on lattice models was on simple cubic
lattices, Skolnick and
others\cite{Skolnick90a,Sikorski90,Skolnick91,Kolinski91,Kolinski92}
have used more complex lattices which are able to more faithfully
represent the structure of actual proteins. Using these lattices they
are able to model real protein structures ({\em e.g.} secondary
structure) and study the dynamics of folding and the formation of
these structures.  However, with increasing complexity it becomes more
difficult to study these models in great detail.

Rather than trying to model real proteins exactly, some have opted for
simpler models which permit a more thorough analysis. Chan and
Dill\cite{Chan91a,Miller92,Chan93a,Chan94} have used a two-dimensional
simple cubic lattice model with two monomer types (a polar monomer,
{\sf P}, and a hydrophobic one, {\sf H}). The potential used models
the hydrophobic interaction and is equal to $-\epsilon$ times the
number of hydrophobic contacts ({\sf HH}). They studied short chains,
which allowed them to do exhaustive enumeration to measure a variety
of properties (both static and dynamic). For dynamics they used both
Monte Carlo\cite{Miller92} and transfer matrix
methods.\cite{Chan93a,Chan94} By using short polymers, they were able
to construct the full transfer matrix (this matrix determines the
probability of one state transforming to another) and use it to solve
exactly for the dynamics of the system. Although their model is
simpler than an actual protein, it has yielded a wealth of interesting
information and provided valuable insight into proteins and
heteropolymers. Their models show a two-phase process similar to that
found in proteins. There is a rapid collapse to compact states,
followed by slower reconfiguring of the chains to the native
structure. Fiebig and Dill~\cite{Fiebig93} show that simple searching
strategies, such as the formation of opportunistic hydrophobic
contacts, can lead to the globally optimal conformation (native
state), suggesting a possible mechanism for folding. Shakhnovich and
others\cite{Shakhnovich91} have studied the folding of random
heteropolymers (the interaction between monomers is picked from a
random distribution) on the three-dimensional simple cube lattice.
They examined 27 monomer polymers using Monte Carlo dynamics and also
found a two-stage collapse process in folding.  They found that by
examining an overlap function, which measures how low-energy
conformations differ, they could distinguish the difference between
foldable and not foldable sequences. From examination of many
different sequences, they conclude that the existence of a pronounced
energy gap between the native state and the remaining conformations
distinguishes good folding
sequences.\cite{Shakhnovich90a,Shakhnovich93} To examine how the
specific form of the interaction affects the dynamics of folding,
Camacho and Thirumalai\cite{Camacho93} looked at two-dimensional
lattice systems.  They studied the kinetics of three different types
of interaction potentials. They found two transition temperatures: a
collapse temperature at which the chain forms a compact structure and
a folding temperature at which the native structure is formed. They
found three stages in the transition from open coil to native
structure.

In this work we will continue using the three-dimensional simple cubic
lattice model. The polymers will be 27 monomers long and consist of
two monomer types. Monte Carlo dynamics will be used to study the
collapse and folding kinetics. The chains are too long for exhaustive
enumeration of all conformations but are short enough to permit
exhaustive enumeration of all maximally compact configurations. This
information will be used to determine the minimum energy structure
(native state) which will allow us to measure the folding time from
extended conformations. We will examine several different sequences
and measure collapse and folding time as a function of temperature and
sequence. One question to be addressed is which kinetic quantities are
sequence dependent and which are sequence independent ({\em
  self-averaging}). In addition, we will examine how the glass
transition affects the ability of a sequence to fold. A major goal is
to define, as precisely as possible, various physically important
quantities. Of particular importance will be the determination of the
important time scales. One problem with Monte Carlo dynamic
simulations is the relation between Monte Carlo steps and physical
time. There is no simple connection; in fact, the precise relation may
depend on the move set.\cite{Chan93a,Chan94} To circumvent this
problem, we will relate Monte Carlo steps to physical time by looking
for the natural time scales in the problem, such as the collapse and
the folding time.  Using these time scales, we will then be able to
define the glass transition temperature ($T_g$) of this model.  In the
past others have speculated that the relation between the folding
temperature ($T_f$) and the glass temperature ($T_g$) would play an
important role in protein folding. Bryngelson and
Wolynes\cite{Bryngelson87,Bryngelson89} have proposed that in order
for a chain to fold, the folding transition must occur before the
glass transition of the system, and the optimal folding temperature
would lie between $T_f$ and $T_g$. Specifically, Wolynes and others
state that to optimize folding potentials for structure prediction,
one should maximize the ratio of the folding temperature to the glass
temperature ($T_f/T_g$).\cite{Sasai90,Goldstein92} To calculate the
glass transition, they used a random energy model-like assumption;
{\em i.e.}, for each given value of the degree of folding, the
energies of the different conformation are independent random
variables. In our work we will give a direct kinetic definition of the
glass temperature that does not rely on this assumption, and show
explicitly that the relative values of $T_g$ and $T_f$ will determine
the folding properties of a given sequence.

\section{MODEL \& METHODS}

The model we used in this work is a three-dimensional lattice polymer.
Monomers that are connected along the chain are constrained to be
nearest-neighbors on the lattice, and only one monomer is allowed per
site. (This is the excluded volume condition.) The chain is then a
self-avoiding walk on the lattice. The polymers are all 27 monomers
long. The maximally compact state is a {$3\!\times\!3\!\times\!3$}
cube (see figure~\ref{fig:cube}). Although it is not feasible to
enumerate all configurations of a 27 monomer chain, it is easy to
enumerate all the compact cubes, of which there are 103346.  If we
choose a potential that favors the formation of contacts, then the
minimum energy conformation will usually be a compact cube. Selecting
such a potential enables us to determine the native structure of a
given sequence by enumeration of the cubes, since for this simple
model the native state is the lowest energy conformation. In addition,
the degeneracy of the lowest energy state can be determined. Since we
are interested in protein-like polymers which have a ``single'' native
state,\cite{Note02} we will choose sequences with a non-degenerate
ground state, {\em i.e.}, those with only one lowest energy
conformation.

We want a potential that will favor compact states and cause the chain
to fold. The dominant force in protein folding is the hydrophobic
effect.\cite{Dill90c} This force is a many-body interaction between
the hydrophobic side chains and the solvent (water). The main effect
is to cause the chain to collapse and create a hydrophobic core. In
our simulations we model this effect by using an attractive potential
to collapse the chain. This potential favors the formation of contacts
between any two monomers. However, we do not want a homopolymer, so
the interaction energy is dependent on whether the two monomers in
contact are of the same type or not.  The
potential is given explicitly by
\begin{equation}
  E = \sum_{{\left\langle
      i,j\right\rangle}\atop{\left|i-j\right|\neq1}} H_{t_i,t_j},
\end{equation}
where the sum is over all nearest neighbor pairs on the lattice,
excluding covalently linked pairs. The type of monomer $i$ is $t_i$
which we will denote with {\sf A} and {\sf B}. $H_{t_i,t_j}$ is the
interaction matrix given by
\begin{equation}
  H_{t_i,t_j}=\bordermatrix{ &{\sf A}&{\sf B}\cr {\sf A}& E_l & E_u\cr
    {\sf B}& E_u & E_l\cr}.
\end{equation}
$E_{l}$ is the energy for a contact between monomers of the same type,
and $E_{u}$ is for contacts between unlike monomers.  To collapse the
chain, we pick both energies to be negative with $E_l<E_u<0$, favoring
contacts between monomers of the same type.

We now need to specify the dynamics of the model. For lattice systems
there is no unique way to do so, and it has been shown that different
move sets may give very different kinetic behavior. In a study of
homopolymer folding kinetics, Chan and Dill\cite{Chan93a,Chan94}
showed that various kinetic quantities, like the collapse time and the
mean first passage time, will depend on the type of moves allowed.
Therefore, we wish to choose a set of moves that will give dynamics
that are as realistic as possible. Care must be taken in analyzing the
results of these simulations. In particular, one must not try to
extract too much detail from these types of simulations. The right
questions need to be asked. For example, we will look at how folding
and collapse time varies with sequence. This is a generic question and
the behavior should be universal to all reasonable move sets. A
question which would be more difficult (or perhaps not even valid) for
this simulation to answer would concern specific details of the
folding pathway, for instance, the role of secondary structure
formation in folding. One would imagine that depending on the move set
used one would find very different answers to this question. To answer
such questions, more realistic models with clearly defined dynamics
are necessary.

The move set used consists of local moves which preserve the covalent
links and keep each lattice site either singly occupied or empty. This
set was developed some time ago to study the dynamics of
polymers.\cite{Verdier62,Hilhorst75,Kremer88} The allowed moves
consist of end moves in which the ends of the chain move to an empty
adjacent site, corner moves which flip a single monomer and crankshaft
moves which move two monomers simultaneously (see
figure~\ref{fig:moves}). Studies involving this move set have shown
that it closely reproduces the relaxation dynamics of the Rouse
model.\cite{Kremer88,Guler83} More complex moves are possible where
more than two monomers are moved simultaneously. They have the
advantage of allowing concerted motion of structural elements (like
helices).  One must take into account the different rates of these
more complex moves; {\em i.e.\/}, a 5 monomer motion should occur more
slowly than the flipping of a single monomer. If this is not taken
into account, the time scales will be distorted since different moves,
all of which can be performed in one iteration of the algorithm, will
have different ``physical'' times. To correct for this, one can assign
a different probability for each of the
moves.\cite{Sikorski90,Skolnick90a} Since our chains are relatively
short, we use only the simple one and two monomer moves. We do not
believe that including the possibility of concerted motions of large
subsections of the polymer will change the answers to the questions
asked here.

There is one important comment to be made about this set of
moves---they are not ergodic. In particular, it is not possible to
reach the configuration in figure~\ref{fig:knot} from an extended
chain.\cite{Madras87} The question is whether the non-ergodicity of
our moves will create a problem. The simple answer to this problem is
that we are interested in the kinetics of folding and as long as the
native state is accessible, there should no problems due to the
existence of inaccessible states. In particular, we can view these
states as being irrelevant in the same way that highly unlikely states
are irrelevant for real proteins. The chain in figure~\ref{fig:knot}
actually forms a knot. We know that real proteins do not have
``tight'' knots,\cite{Schulz79,Note01} and it is highly unlikely that,
in folding, a protein passes through a knotted state.  Strictly
speaking, it is not impossible for a protein to become knotted; it is
just unlikely. Due to the constraints of the lattice, the knotted
state is now inaccessible rather than unlikely; however, its existence
will have no effect on the folding properties. One may argue that
these inaccessible states may still affect thermodynamic calculations.
In particular, what will be the effect of the fact that our moves
restrict us to a ergodic subspace of the full phase (conformation)
space of the system?  That will depend on the relative sizes of the
excluded space. In practice, for small chains, the errors introduced
by the non-ergodicity of this move set is smaller than the statistical
error. Comparison with other ergodic algorithms show no change in the
results.\cite{Kremer88}

A move is made using the Metropolis Monte Carlo
algorithm.\cite{Metropolis53} A monomer is selected at random.  If it
is an end monomer, then one of the neighboring lattice point is also
selected at random. If it is not an end piece, then it can do either a
corner move or a crankshaft move, depending on the position of its
neighbors along the chain. In the case of the crankshaft, the possible
direction is also selected at random. If the move selected would
violate the excluded volume constraint by moving the monomer to an
occupied site, the old configuration is counted once more in averaging
({\em i.e.}, a step is considered to have elapsed), and a new monomer
is picked. If a move is possible, that is, if the lattice site is
empty, then the energy of the new conformation is calculated and
compared to the original energy.  If the energy decreases, then the
move is accepted unconditionally.  If the energy increases, then the
move is accepted with the usual Boltzmann probability:
\begin{equation}
  P = \exp\left[-(E_{\rm new}-E_{\rm old})/T\right],
\end{equation}
where $E_{\rm new}$ and $E_{\rm old}$ are the new and old energies,
respectively, and $T$ is the temperature. Note that we have chosen
units such that the Boltzmann's constant is unity ($k_{\rm B}=1$).
Whether a move is accepted or not, one unit of time (a Monte Carlo
step) is considered to have elapsed.

There is no simple and direct connection between Monte Carlo steps and
the physically relevant times scales of the system. One important
result of this work is that we will determine the mapping between the
physically important time scales such as the folding time and the
computation time scales (Monte Carlo steps). This will be useful in
later works in which we will study the thermodynamics of these systems
where it will be necessary to know how long it takes for systems to
reach equilibrium and explore conformation space. Once again, the
precise connection between physical time (like the folding time) and
Monte Carlo steps will depend on the details of the move set used.
However, we expect that as long as the moves are chosen with care,
that is, one attempts to make it as physically realistic as possible,
then the qualitative features will remain the same. For example, the
behavior of folding time as a function of temperature will look
qualitatively the same although the exact folding time (number of
steps) will vary.

\section{RESULTS AND DISCUSSION}

In this work we studied six sequences, all 27 monomers long. Four were
selected by the following procedure. First a sequence was generated at
random with the appropriate ratio of monomer types. We then enumerated
all cubes calculating the number of minimum energy states. Sequences
with degenerate minimum energy states were rejected. From the
remaining non-degenerate sequences we picked four which had a spread
in energy of the native states ($-82$ to $-76$). One sequence was
obtained by changing a single monomer in one of the original four;
{\em i.e.\/}, it is a single-site mutation, which lowered the ground
state energy, from $-82$ to $-84$. The last sequence was taken from a
paper by Shakhnovich\cite{Shakhnovich93} which gave a method for
finding optimal sequences; it also has a ground state energy of $-84$.
Table~\ref{tab:seq} shows the various sequences along with some data
for each.  All six sequences have the same ratio of monomer types,
roughly 50:50 (14:13, actually). For these simulations the value for
the contact energies are $-3$ for monomers of the same type and $-1$
for unlike monomers.  Again, the units used are such that $k_{\!B}=1$.
The maximally compact conformation of a 27 monomer chain is the
{$3\!\times\!3\!\times\!3$} cube. There are 28 non-covalent or {\em
  topological} contacts in this cube, so the lowest possible energy is
$-84$. Two of the sequences have this energy for their minimum
conformation.  This corresponds to a completely ``unfrustrated''
ground state.  By ``unfrustrated,'' we mean that the ground state has
no topological contacts between monomers of different types. The other
three sequences have ground state energies higher than $-84$ and,
consequently, their ground states have at least one bad topological
contact and consequently are frustrated energetically.

Using the Monte Carlo moves described above, we simulated the folding
of these sequences and examined how the folding behavior depends on
temperature and differs from sequence to sequence. For each sequence
we started with a random, completely unfolded, initial conformation.
Here, completely unfolded means that there are no contacts between any
of the monomers. A temperature was selected, and the sequence was
allowed to fold. Once the sequence found its folded state (which we
detected by monitoring the energy), we stopped the simulations. The
simulations were also stopped if the sequence did not find its folded
state within ${\tau_{\!\rm max}}$ steps. For the simulations in this
work, ${\tau_{\!\rm max}} = 1.08\times10^{9}$ Monte Carlo steps.  This
maximum time was picked both so that it was longer than the typical
folding time of most sequences and to minimize the actual computer
time used in the simulations. It is important to realize that any
times longer than ${\tau_{\!\rm max}}$ are undefined. Ideally, we
would want to pick ${\tau_{\!\rm max}}$ to be longer than any
interesting and relevant physical or biological time scale. Since
there is no simple connection between Monte Carlo steps and physical
time, we cannot directly determine ${\tau_{\!\rm max}}$. We chose a
first value for ${\tau_{\!\rm max}}$ which seemed reasonable and then
made sure that it was much longer than the folding time for the
various sequences.

At each temperature we ran many simulations, each with a different
random initial condition (always unfolded). We then calculated an
average folding time (${\tau_{\!f}}$) from these runs. This time is
the mean first passage time from the set of unfolded initial states to
the folded state.  Figure~\ref{fig:ftime1} shows ${\tau_{\!f}}$ as a
function of temperature. Once again, the units of temperature and
energy have been chosen so that $k_{\!B}=1$.  We ran anywhere from 10
to 600 simulations at each temperature and calculated the average
folding time. If the folded state was not found within ${\tau_{\!\rm
    max}}$ steps, we averaged in ${\tau_{\!\rm max}}$ for that run.
The error bars are the standard deviation of the mean given by
$\sigma/\sqrt{N}$, where $\sigma$ is the standard deviation of the
distribution of folding times and $N$ is the number of runs at that
temperature.  It is important to note that since we average in
${\tau_{\!\rm max}}$ when the chain does not fold, the error bars are
not as meaningful at temperatures where the folding time approaches
${\tau_{\!\rm max}}$ and may be much larger at these temperatures. In
particular, at high and low temperature the points equal ${\tau_{\!\rm
    max}}$ with zero error. That is simply due to the fact that at
those temperatures the simulations never found the folded state.
Figure~\ref{fig:fracfold} shows the fraction of times the folded state
was found as a function of temperature.  It has a maximum plateau over
the same temperature range that ${\tau_{\!f}}$ has a minimum plateau.
At temperatures where the chain folds rapidly, it also finds its
native state 100\% of the time.  At temperatures where the simulations
did not find the folded state all the time, the ${\tau_{\!f}}$ shown
in figure~\ref{fig:ftime1} is a lower bound to the actual mean first
passage time.  Figure~\ref{fig:ftdist} shows the distribution of
folding times for three temperatures. At the temperature of fastest
folding ($T=1.58$) the distribution of times is narrowest. For
temperatures above and below this the distribution becomes quite
broad.  All three histograms appear roughly Poissonian.  The standard
deviations are approximately equal to the means.

We observe three different temperature regions, similar to those found
in two-dimensional lattice simulations.\cite{Miller92,Chan94} Above a
temperature of $\sim3$ and below $\sim.65$, the chains did not fold
within ${\tau_{\!\rm max}}$ steps. Between these temperatures the
folding time drops rapidly to approximately
$2\times10^7$--$5\times10^7$. The fraction of runs that find the
folded state increases sharply from 0 to 1 in this temperature range.
In the next plot (figure~\ref{fig:comptimes1}) the folding time is
plotted along with the chain compaction time. The compaction time is
simply the number of steps it takes for an unfolded state to reach a
maximally compact cube. In addition, we also show a time to reach a
nearly compact state, which we define to be a conformation with 25
(out of 28) contacts. The behavior of the compaction time as a
function of temperature is similar to that of the folding time, but
chains compact much faster than they fold. This behavior is similar to
what is believed to occur in real proteins in which the chain first
folds rapidly to a compact state and then rearranges itself to the
native structure.

Above a temperature of approximately $5$, the compaction time
approaches ${\tau_{\!\rm max}}$. Above this temperature the
free-energy is dominated by the entropic term which favors non compact
conformations which are far more numerous than compact ones.  In the
range from $5$--$2$, we observe the following interesting behavior.
The chains compact fairly rapidly, but the folding time is still quite
long. In particular, at about $T\approx3$ the chains compacted easily
but never folded within ${\tau_{\!\rm max}}$ steps. We can draw a
parallel between this state and the molten globule state of
proteins.\cite{Ptitsyn92} At this temperature range, the temperature
is low enough so that the potential, which favors contacts, drives the
chain to a compact conformation, but the temperature is still too high
for the potential to drive the chain to the native state.  In this
range we can imagine the chain is fluctuating about various compact
states, ``randomly'' searching for the native state. This is like the
often-discussed Levinthal paradox in which it is argued that a protein
could not find its folded structure by random search.  If the
temperature is high enough there is no strong driving force that
favors the native state.  When the temperature is low enough, the
chain is no longer randomly searching compact conformations but is
driven to the folded state.  This is, of course, the well-known
resolution to the paradox.  At the appropriate temperatures proteins
do not randomly search for their native state but are directed to it
by the shape of the free energy surface.

At still lower temperatures both the folding and compaction time start
to increase again. At temperatures slightly less than 1 the folding
time reaches ${\tau_{\!\rm max}}$ again and at a temperature of
roughly $.63$ the compaction time approaches ${\tau_{\!\rm max}}$. At
low temperatures the system is beginning to slow down, kinetically,
and is now getting trapped in local meta-stable states. Even though we
expect, at these low temperatures, the free energy to have a very
pronounced minimum at the native state, the system is unable to reach
it within a reasonable time. This region is often referred to as the
glass phase. We can define a temperature at which the system undergoes
a glass transition characterized by the slowing down of various times,
such as the folding and compaction times. The autocorrelation time of
the system would also increase in this temperature region, indicating
that the chain was locally trapped. We define the glass transition
temperature ($T_g$) as the temperature at which the folding time is
half way between ${\tau_{\!\rm max}}$ and ${\tau_{\!\rm min}}$ (where
${\tau_{\!\rm min}}$ is the fastest folding time observed).  Using
this definition we get $T_g\approx1$.  Note that the definition of
glass temperature is not the usual thermodynamic definition of
temperature. It is not determined by the inverse of the derivative of
entropy with respect to energy ($1/T_g = \partial S/\partial E$) at
the point where the entropy ``vanishes.''\cite{Derrida81} One
difficulty with this definition is its relation to the kinetics of the
system.  The idea is that as a system is taken out of equilibrium,
then time it takes to relax back will increase as the temperature gets
closer to $T_g$. To avoid this kind of assumption, we have given a
kinetic definition for $T_g$ in which we will explicitly look for a
slowing down of the system.  We would expect the precise value of
$T_g$ to depend on the moves used and therefore we will not focus on
the exact value but on the relative value. In particular later we will
compare $T_g$ to another important temperature, the folding
temperature ($T_f$), and we will discover a key relation between the
two.  Additionally $T_g$ depends on the value of ${\tau_{\!\rm max}}$;
that is, it will depend on how long we run our simulations.  This is a
subtle but extremely important fact to remember when studying finite
sized systems. When talking about glass-like behavior of a finite
system the notion of glass-like depends on the time scale you are
looking at. If you wait long enough the chain will always find the
native state. To speak of a physically meaningful glass transition one
must define the physical time scales of interest.  The time scales of
importance here are related to the minimum folding time of a good
folding sequence. We want to examine our system on a time scale that
is reasonable greater then the minimum folding time. For our simple
system there is no obvious greater time to pick. For real proteins the
life time of the organism would be a reasonable choice, since proteins
need to fold on a time scale much shorter than this to be useful. We
picked a time that was roughly two orders of magnitude greater than
the folding time.  Since this is somewhat arbitrary, we investigated
how $T_g$ changed as ${\tau_{\!\rm max}}$ is varied.
Figure~\ref{fig:tg} shows the results of several simulations in which
${\tau_{\!\rm max}}$ was varied from one fourth the usual value
($1.08\times10^{9}$) to almost twice this value. We see that although
$T_g$ decreases with increasing ${\tau_{\!\rm max}}$ it does so quite
slowly. The difference between the last two is only about 4\%;
therefore $T_g$ is not too sensitive to the precise value of
${\tau_{\!\rm max}}$.

The glass temperature just defined is related to the folding of the
chains. One can also define a glass temperature that has to do with
the slowdown of compaction. This would be the temperature at which the
time it takes the chain to form a cube (28 contacts) is half way
between the maximum and minimum times. Call this temperature
$T_g(28)$. Examining figure~\ref{fig:comptimes1} we see that $T_g(28)$
is less than $T_g$. (It is approximately equal to $0.7$.)  One could
also consider $T_g(25)$ the glass temperature for forming 25 contacts
(which is lower still). In general the transition temperature will be
a function of some parameter, $\rho$, which is a measure of the
compactness of the chain and/or similarity to the native state.
Bryngelson and Wolynes\cite{Bryngelson89} first calculated $T_g(\rho)$
in their random energy model.

The two regions in which the chain fails to fold are qualitatively
very different. At high temperatures the energy differences between
conformations becomes negligible so all conformations have roughly
equal probabilities. The chain is randomly exploring the conformation
space.  It takes a long time to find the native state by random search
due to the vast number of conformations. The free energy is dominated
by the entropic term so the native state is no longer the global
minimum. At low temperatures the energy differences between states
becomes important and the folded state is the global minimum free
energy. The problem now is that the barriers between states are too
high and at low temperatures there is a very small probability for
crossing them.  For compact conformations many moves will involve the
breaking of contacts which at low temperatures becomes unlikely. In
particular, moves that break more than one contact are much less
probable that those that break only one. Instead of a random search
the chain is now forced in to a very narrow kinetic pathway consisting
of those steps with very small free energy barriers. The chain gets
trapped in the many local minima.

If we were willing to wait long enough the system would eventually
fold. Since our system is finite, the system always has a finite
non-zero probability to find the native state.  The same is true for
the high-temperature case: if we wait long enough the chain will
eventually find the folded state.  However, one must remember that at
those temperatures the folded state is not the free energy minimum and
is therefore not stable.  For example, consider the following two
temperatures: $2.24$ and $1.12$.  The folding time for these two
temperatures is roughly the same.  At the higher temperature (as we
will see shortly) the chain spends almost zero time in the folded
state (less than 0.04\%). At the lower temperature the chain spends
roughly 77\% of the time in the folded state.  When we speak of
folding time, this is simply the time it takes the chain to find the
native conformation.  There is another important factor here: namely,
is the folded state stable thermodynamically? We will address this
issue at the end of this paper, where we see that it is not enough
that a chain find its native state in a short time, but it must do so
at temperatures where the native state is thermodynamically stable.

Let us return to the question of how long is too long to wait for a
sequence to fold. Too long is in general determined by other time
scales in the system. For proteins, there are a number of biologically
relevant time scales, the lifespan of the organism for example.
Proteins that do not fold fast enough on this time scale can be
considered not to fold at all. Since we are studying a simple
artificial system there is no {\em a priori} time scale to pick, other
than limits on the simulation (computer) time. One of the problems
with Monte Carlo dynamics is that there is no easy way to
``calibrate'' them, that is to make a connection between Monte Carlo
steps and ``real'' physical or biological time.  What we have done
here is to define the relevant time scale as the folding time (or the
compaction time) and make sure we ran simulations for long enough that
we could see the the variation of folding time as a function of
temperature.

We have examined the folding (and compaction) time as a function of
temperature for one sequence. We now would like to see how this
function varies from sequence to sequence. Figure~\ref{fig:comptimes}
show a plot of the folding time and compaction time versus temperature
for several sequences. From this figure we notice two very interesting
features of this model. First the compaction time is sequence
independent. All six sequences have roughly the same compaction time
for temperatures above $T_g$. In contrast the folding time is highly
sequence dependent. At a temperature of roughly $1.6$ there is a
difference of nearly an order of magnitude between the fastest and
slowest folding sequence. The folding time is also roughly correlated
with the energy of the folded state. The lower the energy, the faster
the folding time. However, the relation between the energy of the
native state and the folding time is not a simple one. For example the
two sequences with the lowest energy folded states (seq.\ $002$ and
$004$, see table~\ref{tab:seq}) have different folding times. The
difference is slight but sequence $004$ has a consistently faster
folding time at all temperatures. Sequence $005$, which is a single
monomer mutation of $004$, has a higher ground state energy ($-82$),
but its folding times are very close to those of the lower energy
sequence $002$. There also appears to be a fairly large difference
between the sequences that have energies below $-81$ and those that
have energies above.

Note that the collapse time is always much faster than the folding
time for all sequences. Even sequences that fold slowly collapse as
rapidly as the fast-folding ones. This sequence-independent property
of the collapse time is often referred to as a {\em self-averaging}
property; it does not depend on the specifics of the sequence but
rather on the general character of the ensemble of sequences. It is
important to remember here that we are choosing a restricted ensemble
of sequences though, namely the sub-set of sequences with a particular
ratio of monomer types (a ratio of 14:13).  Sequences that contain a
different composition of monomers may have different collapse times
than the sequences used here.  The folding time is not self-averaging;
{\em i.e.}, it depends on the sequence. So we can view the kinetics of
folding as a two-stage process. The first involves a rapid collapse of
the polymer. The nature of this collapse is sequence independent. We
can picture the polymer in this collapsed state as fluctuating about
various compact cube states. This picture has been advanced previously
by others.\cite{Leopold92} The next step is a medium-to-slow event in
which the polymer searches for the minimum energy state among the
compact states. The time it takes for the polymer to find its minimum
state depends on the specifics of the sequence. The two-phase collapse
with two distinct time scales has also been observed for real
proteins.\cite{Kuwajima89,Kuwajima92,Baldwin93} The first phase is a
rapid collapse in which a hydrophobic core is formed.  We should
expect this collapse to be independent of the specific sequence,
depending on the ratio of hydrophobic to hydrophilic monomers. This
collapsed state then undergoes rearrangement to the folded structure
of the specific sequence. The collapse time below the glass
temperature loses its self-averaging property. Examining
figure~\ref{fig:comptimes} we see that below $T_g\approx1$ the
collapse time is no longer sequence independent. In the glass region
we would expect the kinetics to depend on the details of the energy
surfaces and these details will be sequence dependence. This is
expected of a system exhibiting glassy behavior. Note how this
contrasts with the high temperature limit, where the collapse times
remain sequence independent even as they approach ${\tau_{\!\rm
    max}}$.

At this point one may be tempted to conclude that we have two types of
sequences: fast folders and slow folders. However this is not the
case. In reality what we have are sequences that fold and sequences
that do not. In order to see this we need to look at the
thermodynamics of these systems. In particular we need to look at the
thermodynamic stability of the native state as a function of
temperature. To do so, we performed a series of thermodynamic runs
using the same Monte Carlo algorithm described above.  The system was
equilibrated by first running it for 100 million steps, which is on
the order of the folding time for most of the sequences. Care must be
taken at low temperatures since near the glass transition the system
will slow down; {\em i.e.}, the auto-correlation time will diverge. We
looked at temperatures above $T_g$ to avoid this problem. We calculate
the
following thermodynamic quantity:
\begin{equation}
  P_{\rm nat}(T) = \frac{e^{-E_{\rm nat}/T}}{Z},
\end{equation}
where $E_{\rm nat}$ is the energy of the native state and $Z$ is the
partition function. This quantity is the probability that the system
is in the native state; that is, it is folded. We define the folding
temperature as the temperature at which $P_{\rm nat}(T_f)=0.5$; that
is when the folded state is half occupied. Note that $P_{\rm
  nat}(T)>0.5$ is a sufficient condition that the native state be the
global minimum of the free energy. Four sequences were used, each with
a different minimum energy ranging from $-84$ to $-76$.  Several
simulations were run at a number of different temperatures.
Figure~\ref{fig:thermo} shows the results. In addition to running
simulations at different temperatures, we also used the Monte Carlo
histogram method\cite{Ferrenberg88} to calculate the function $P_{\rm
  nat}(T)$ at temperatures other then the simulation temperature.
Using histograms collected from simulations run at $T=1.58$, we were
able to calculate $P_{\rm nat}$ for all temperatures, extrapolating
into the glass region. These lines are plotted along with the points
calculated from the standard Monte Carlo runs.

The folding temperature, $T_f$, varies with the value of the
minimum-energy state. The lower the minimum energy the higher the
folding temperature. More importantly we see that the two lower energy
sequences have folding temperatures above the glass temperature,
$T_g$, while the others have $T_f < T_g$. At $T_g$ the lowest energy
sequence (which also folds the fastest) is 90\% in the native state.
The highest energy sequence has a native state population of only
15\%.  At temperatures at which the folded state of the high-energy
sequence is thermodynamically stable this state is not kinetically
accessible.  Therefore we would say this sequence does not fold. In
order for a sequence to be foldable it must meet two conditions.
First, it must have a reasonably fast folding time, where by
reasonable we mean on the relevant (biological) time scales, and
second, the folding temperature must be above the glass temperature.
The analogous situation would be a polypeptide which had a folding
temperature below the freezing point of the solvent. Such a protein
would not be considered foldable.

\section{CONCLUSIONS}

Using a simple lattice model and Monte Carlo dynamics we have studied
the kinetics (and some thermodynamics) of protein-like heteropolymer
folding. Our results agree with previous works on other simple models
and also match some of the properties of the folding real proteins. We
find that our models display a two-stage folding behavior. First there
is a rapid collapse to a compact state, followed by a slower stage in
which the collapsed state rearranges itself to the native structure.
We find that the folding time has a minimum plateau at intermediate
temperatures and diverges at both high and low temperatures. The same
is true for the collapse time. In this work we have examined the
folding behavior as a function sequence and have discovered several
interesting results. The collapse time and the glass temperature are
both sequence independent (self-averaging) quantities.  The folding
time and temperature are both sequence dependent. The folding time
correlates approximately with the energy of the native state: the
lower this energy the faster the chain folds. This is consistent with
the results found by Shakhnovich\cite{Shakhnovich90a,Shakhnovich93}
that the larger the energy gap of the native state the better the
sequence folds. We did not measure the gap, since there is no clear or
simple definition of the gap in our system. Another way to view this
result is that sequences with unfrustrated native states (native
states with no bad contacts) fold best; {\em i.e.}, we want to
minimize energetic frustration of the ground state. However, we expect
that this may be a property of these simple systems and that in more
complex systems other forms of frustration (geometric or energetic
frustration of conformations other than the native state) may play an
important role.  One would then expect that systems with reduced
frustration should give rise to a large number of conformations that
are rapidly connected kinetically to the native state (rapid compared
to the folding time) or, as first proposed by Leopold and
others,\cite{Leopold92} a ``dominant folding funnel.''

An important point we have tried to stress is the issue of time
scales, in particular the relevant physical time scales for this
system and for protein folding in general. We note that there was no
simple way to connect the computation time (Monte Carlo steps) to
physical time. Rather than attempt to do so, we simply ran our
simulations for a reasonable number of steps and then observed the
folding time for the system. It is this folding time that now becomes
the key time scale. For example, when we say a sequence does not fold
what we mean is that is does not fold within a time that is over an
order of magnitude greater than the folding time for the fast
sequences. Since we are looking at finite systems we know that they
will all fold given enough time. What is important is whether they
fold in a reasonable time where reasonable is the folding time for the
faster sequences. For real proteins, this time scale would be some
suitable biological time.

By examining the behavior of folding time versus temperature we
defined the glass transition temperature of this system. Below this
temperature the kinetics slow down, causing the folding time to
increase rapidly. Also the collapse times lose their self-averaging
property and are now dependent on sequence. Most importantly we
observed that for the slow-folding sequences the folding temperature
(the temperature at which the native state is half populated) is below
the glass temperature. This indicates that these sequences will never
fold since at temperatures where the native state is thermodynamically
stable it is kinetically inaccessible. Good folding sequences have
$T_f$ greater than $T_g$. It has been suggested by
others\cite{Sasai90,Goldstein92} that a good design principle for
optimizing folding would be to maximize the ratio $T_f/T_g$. We
observe this result explicitly in our simulations.

Perhaps the most interesting observation is that even simple systems
such as these display a wide variety of complex and intriguing
properties, many of which are shared by real proteins. This is
particularly compelling in that one can much more easily study these
simple systems and understand their behavior in great detail. By
examining slightly more complex models we hope to understand how much
of protein behavior is unique to proteins and how much is shared by
the general class of heteropolymer systems. Hopefully, much of the
apparent complexity of proteins will be understandable in the context
of simpler model systems.

\acknowledgments

We would like to gratefully acknowledge the computational assistance
of A.~Schweitzer. We also thank S.~Skourtis, P.~G.~Wolynes, and
K.~Dill for helpful discussions. J.~N.~O. is a Beckman Young
Investigator. This work has funded by the Arnold and Mabel Beckman
Foundation and by the National Science Foundation (Grant No.\
MCB-9018768).  J.~N.~O. is in residence at the Instituto de
F\'{\i}sica e Qu\'{\i}mica de S\~ao Carlos, Universidade de S\~ao
Paulo, S\~ao Carlos, SP, Brazil during part of the summers.


\newpage\widetext\tightenlines

\section*{TABLES}

\begin{table}[hb]
\begin{tabular}{l|c|r|r|r|r}
\hline\hline
  Run & Sequence & $E_{\rm min}$ & ${\tau_{\!\rm min}}$ & $T_g$ & $T_f$ \\
\hline
002 & ABABBBBBABBABABAAABBAAAAAAB & -84 & $2.0\times10^7$ & 1.00 &
  1.285(15) \\ \hline
004 & AABAABAABBABAAABABBABABABBB & -84 & $1.6\times10^7$ & 0.96 &
  1.26(1) \\ \hline
005 & AABAABAABBABBAABABBABABABBB & -82 & $2.3\times10^7$ & 0.98 &
  1.15(2) \\ \hline
006 & AABABBABAABBABAAAABABAABBBB & -80 & $5.2\times10^7$ & 1.07 &
  0.95(6)\\ \hline
007 & ABBABBABABABAABABABABBBABAA & -80 & $9.3\times10^7$ & 1.09 &
  0.93(5)\\ \hline
013 & ABBBABBABAABBBAAABBABAABABA & -76 & $9.7\times10^7$ & 1.01 &
0.83(5)\\ \hline\hline
\end{tabular}
\bigskip
\caption{The various sequences used in
  this paper. The last four (005, 006, 007, 013) were generated at
  random. Sequence 002 was optimized by
  Shakhnovich.\protect\cite{Shakhnovich93} Sequence 004 is a single
  monomer mutation of 005 ($B_{13}\rightarrow A$). Both 002 and 004
  have the lowest energies possible for the potential used and have
  native states that are completely unfrustrated. ${\tau_{\!\rm min}}$
  is the fastest folding time for each sequence. $T_g$ is the glass
  transition temperature (calculated with a ${\tau_{\!\rm
      max}}=1.08\times10^9$).  $T_f$ is the folding temperature
  calculated using the Monte Carlo histogram method. The numbers in
  parenthesis indicate the uncertainty of the last digit.}
\label{tab:seq}
\end{table}

\newpage

\newcounter{fig}
\renewcommand{\caption}[1]{{\noindent Fig.~\arabic{fig}. #1}}
\renewenvironment{figure}{\refstepcounter{fig}\begin{singlespace}}{\end{singlespace}\bigskip}

\section*{FIGURES}

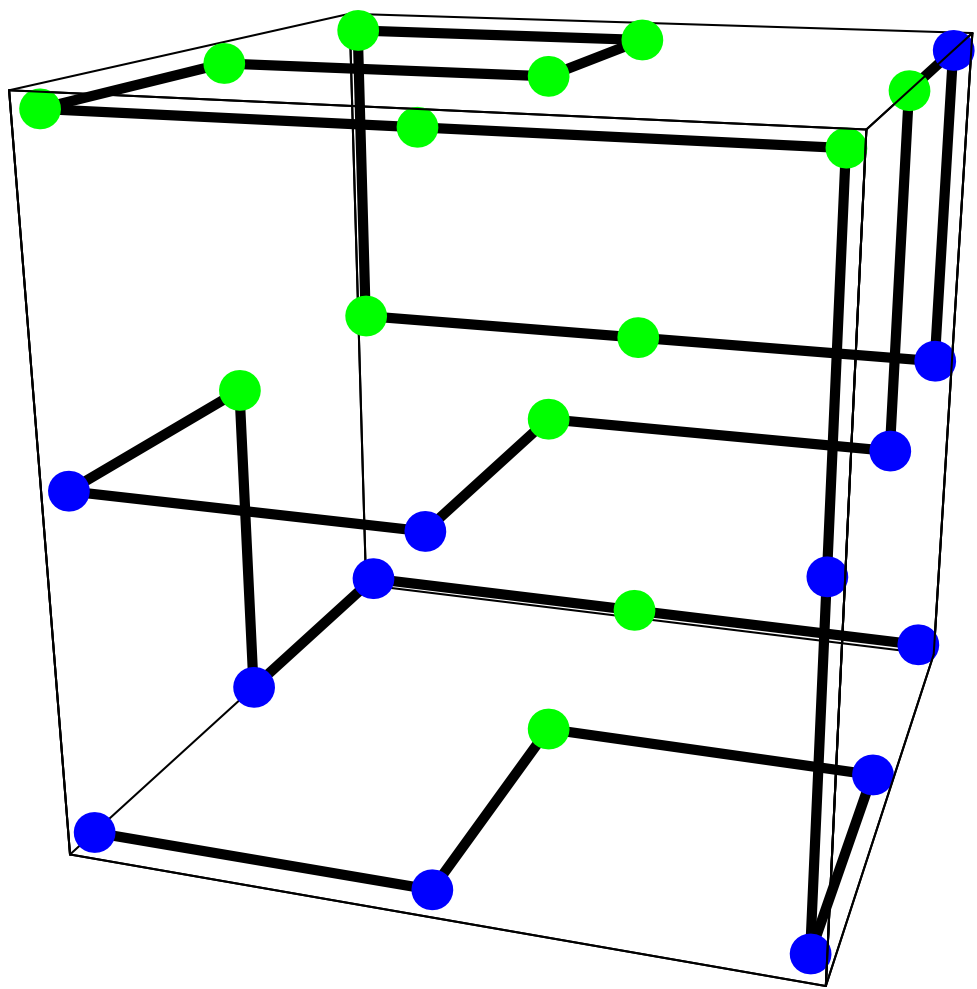
\begin{figure}
\caption{An example 27 length polymer on a three dimensional simple
  cubic lattice. The conformation is a maximally compact cube. The
  light and dark spheres represent the two different types of
  monomers.  The sequence shown here is 013 and the conformation shown
  is the native (minimum energy) state.}
\label{fig:cube}
\end{figure}

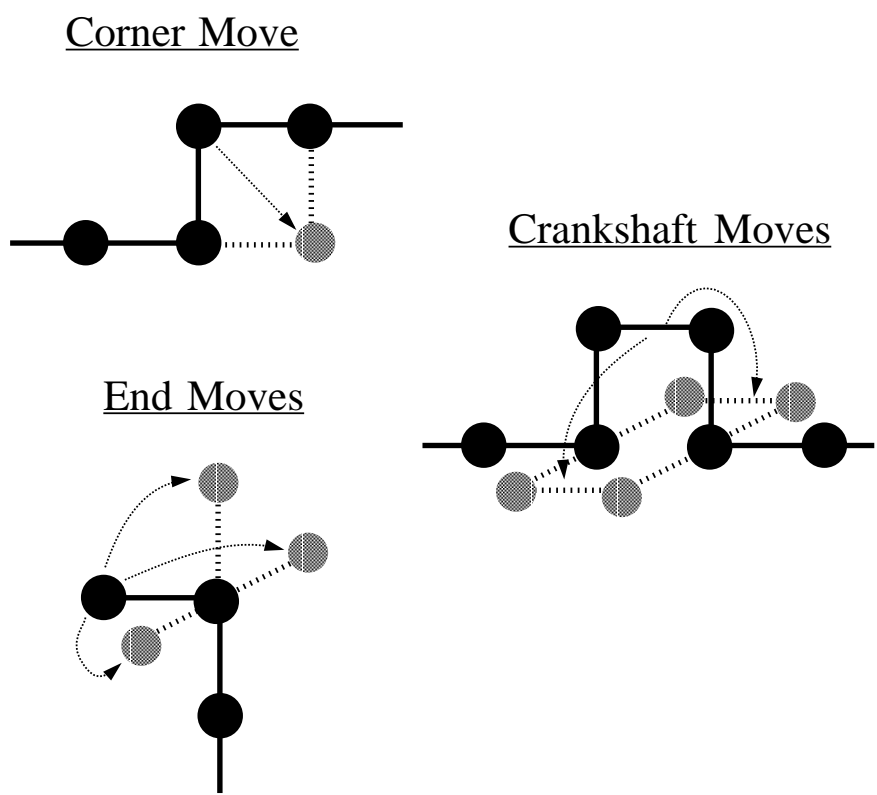
\begin{figure}
\caption{The three types of moves used in the dynamics simulations.
  The light circles represent the possible lattice points a given
  monomer can move to provided that that point is not occupied. In the
  case of the end and crankshaft moves one of the possible moves is
  picked at random. Note that the corner and crankshaft moves are
  exclusive: a non-end monomer can only make one or the other
  depending on the position of its neighbors along the chain.}
\label{fig:moves}
\end{figure}

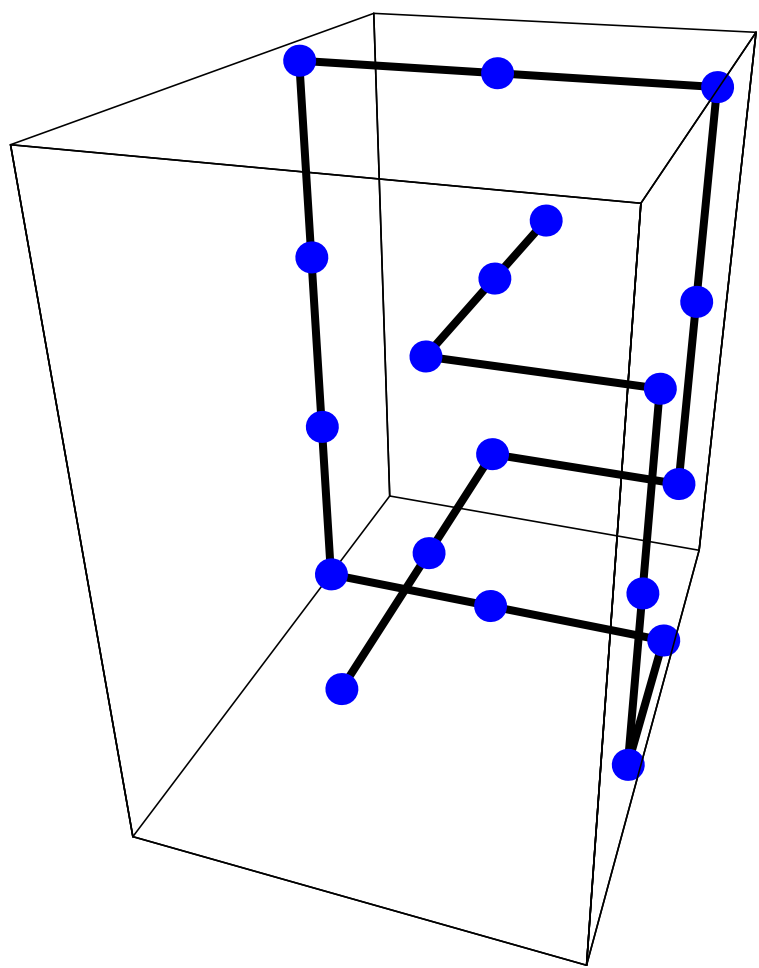
\begin{figure}
\caption{A knotted conformation\protect\cite{Madras87}
  which can not be un-knotted by the moves shown in
  figure~\protect\ref{fig:moves}; consequently, from an unfolded
  conformation it is unreachable using the same moves. Hence the
  moves are not ergodic. As long as conformations like this are not
  the native state they will pose no problem. What is important is the
  the native state is accessible.}
\label{fig:knot}
\end{figure}

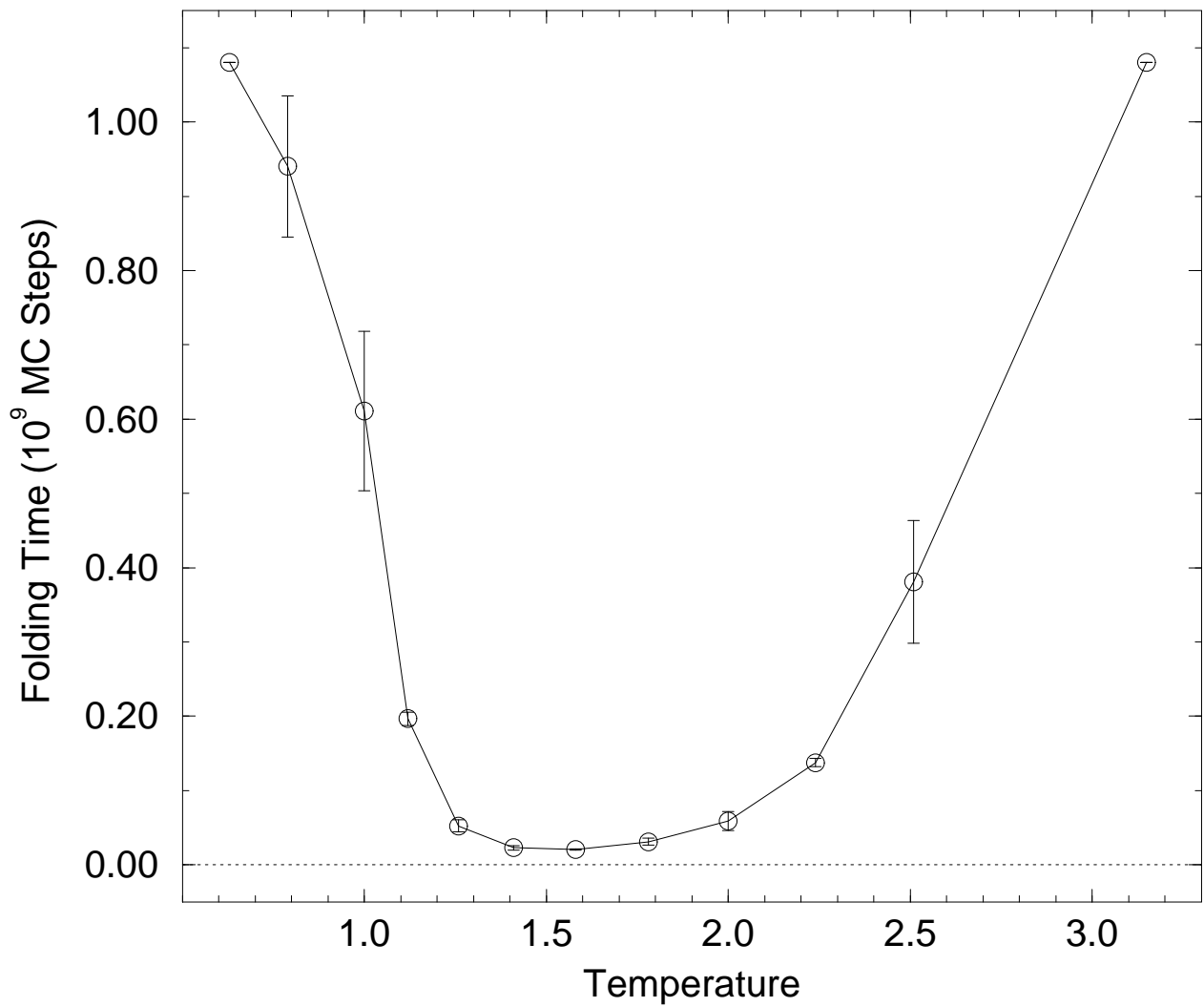
\begin{figure}
\caption{Mean folding times versus temperature for one sequence (sequence
  002). Note both axes are linear scale and the time is in billions of
  Monte Carlo steps. The error bars are the {\em standard deviation of
    the mean}: that is, they are equal to the standard deviation of
  the folding-time distribution at a given temperature divided by the
  square root of the number of runs at that temperature.}
\label{fig:ftime1}
\end{figure}

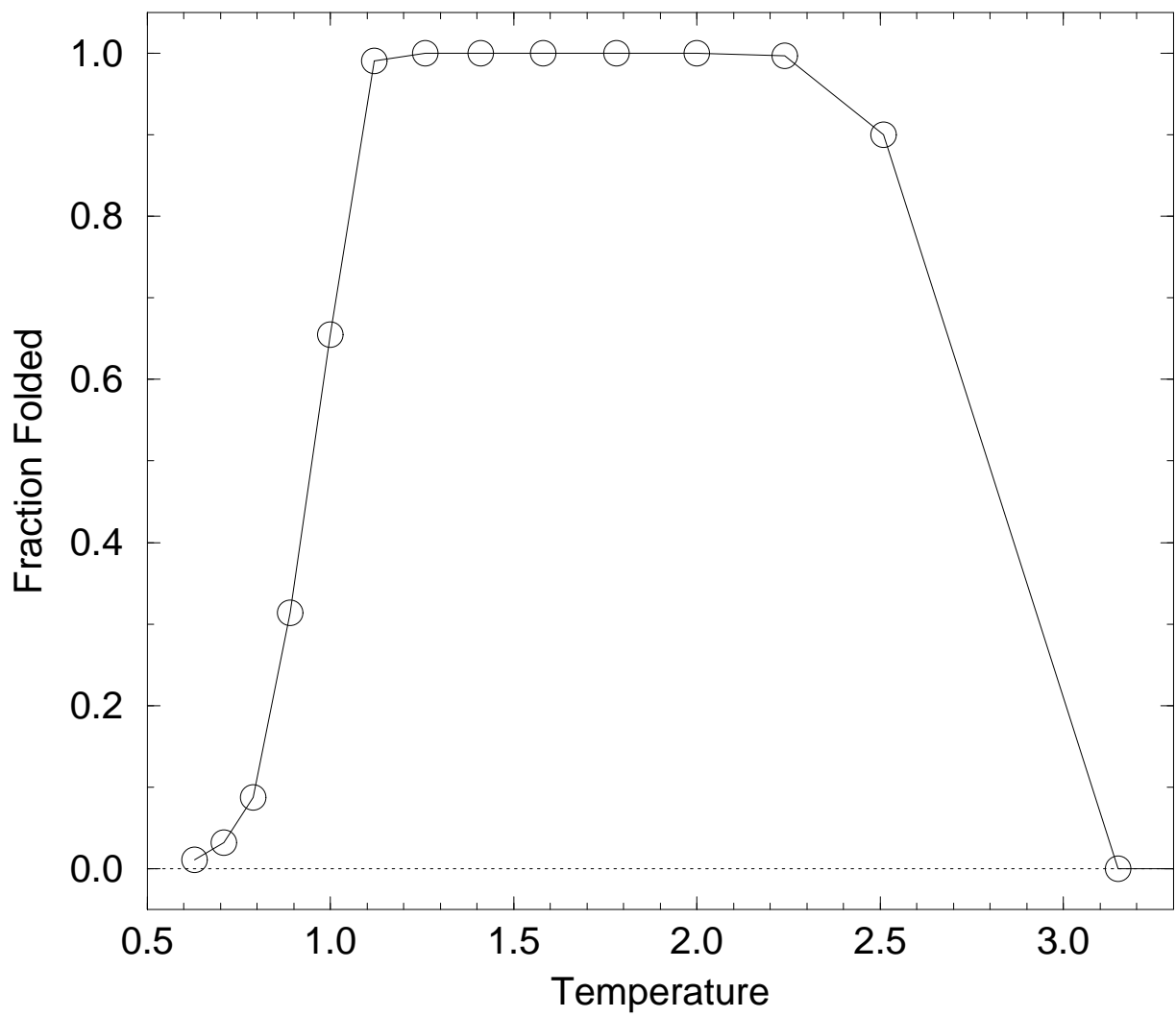
\begin{figure}
\caption{Fraction of times that sequence 002 folded as a function of
  temperature. Note that the plateau at which the chains fold 100\% of
  the time corresponds to the minimum folding time plateau in
  figure~\protect\ref{fig:ftime1}.}
\label{fig:fracfold}
\end{figure}

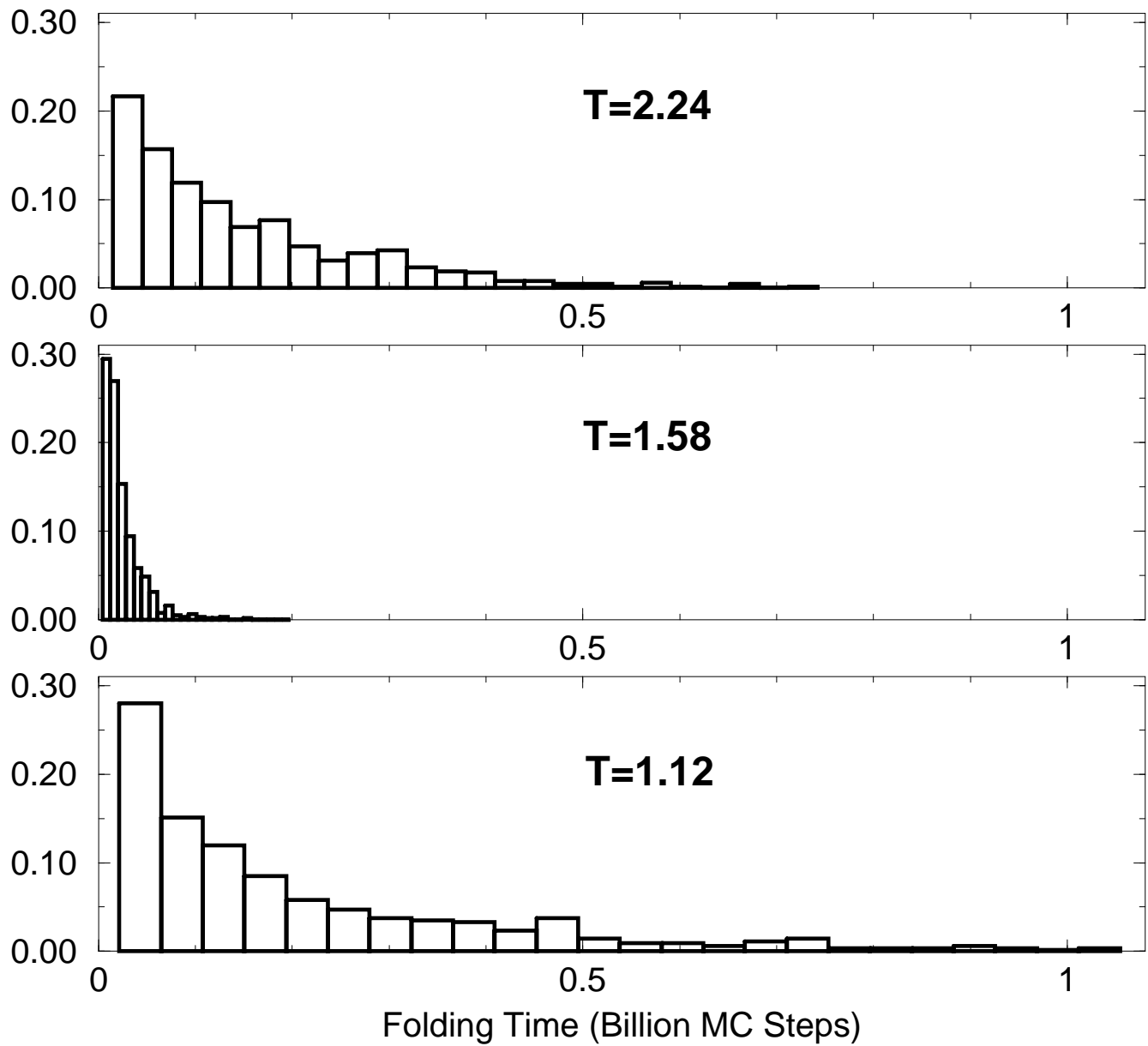
\begin{figure}
\caption{Histogram of folding times for sequence 002 at three
  different temperatures. The histogram values have been normalized so
  the sum over all bins equals one. The minimum folding time
  temperature (1.58) is shown along with distributions above and below
  this temperature. The distributions are roughly Poissonian, the
  standard deviation being approximately equal to the mean.}
\label{fig:ftdist}
\end{figure}

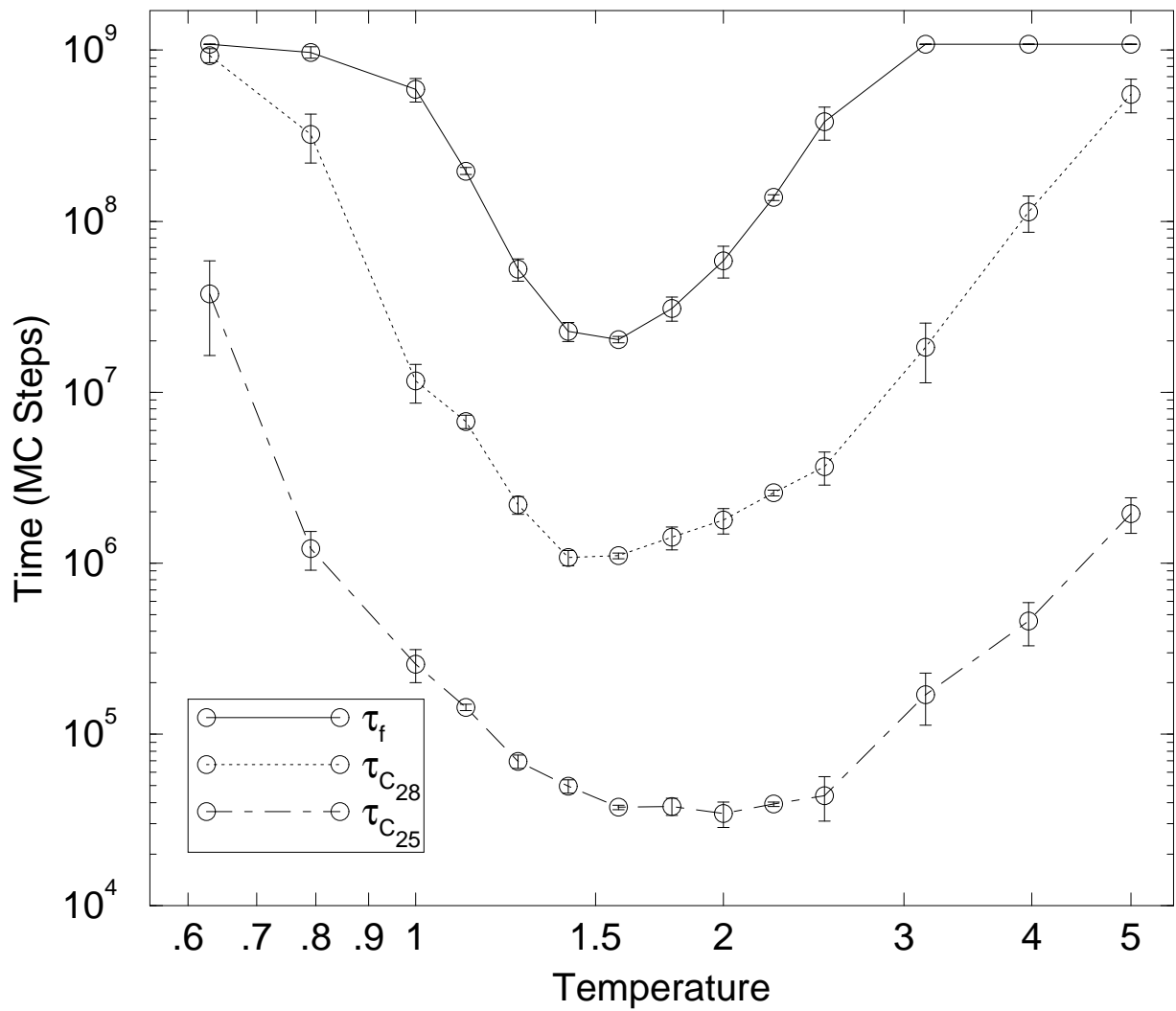
\begin{figure}
\caption{Folding and Collapse times versus temperature for sequence
  002. Note in this figure both axes are log scale and the time is in
  Monte Carlo steps. The solid line is the mean folding time,
  ${\tau_{\!f}}$ (same as figure~\protect\ref{fig:ftime1}, but now on
  a log scale).  The dotted line, ${\tau_{\!c_{28}}}$, is the mean
  compaction time to any cube.  The last line, ${\tau_{\!c_{25}}}$, is
  the mean compaction time to a partially compact conformation with 25
  (out of 28) contacts. Error bars are the standard deviation of the
  mean.}
\label{fig:comptimes1}
\end{figure}

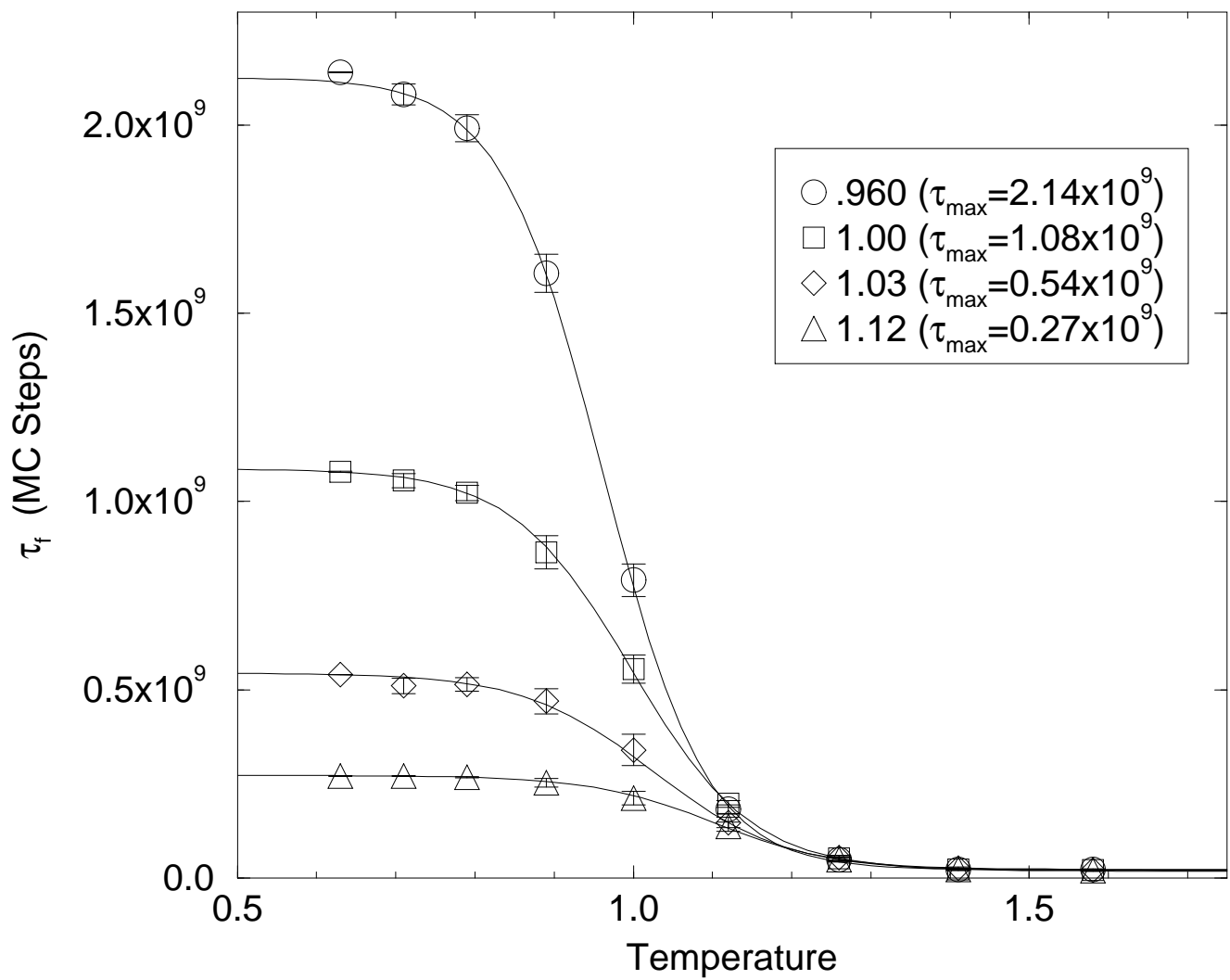
\begin{figure}
\caption{Folding time as a function of temperature at several
  different ${\tau_{\!\rm max}}$.  The glass temperature is the point
  at which the folding time is halfway between ${\tau_{\!\rm max}}$
  and ${\tau_{\!\rm min}}$. The legend shows the glass temperature for
  each value of ${\tau_{\!\rm max}}$.}
\label{fig:tg}
\end{figure}

\begin{figure}
\caption{Folding and collapse times versus temperature for all
  sequences, plotted on a log-log scale. The top set of solid lines
  are the folding times, the middle set of dotted lines are the times
  to compact to a cube and the bottom set are the times to compact to
  partially compact (25 contacts) conformation. The legend shows the
  energy of the native states.}
\label{fig:comptimes}
\end{figure}

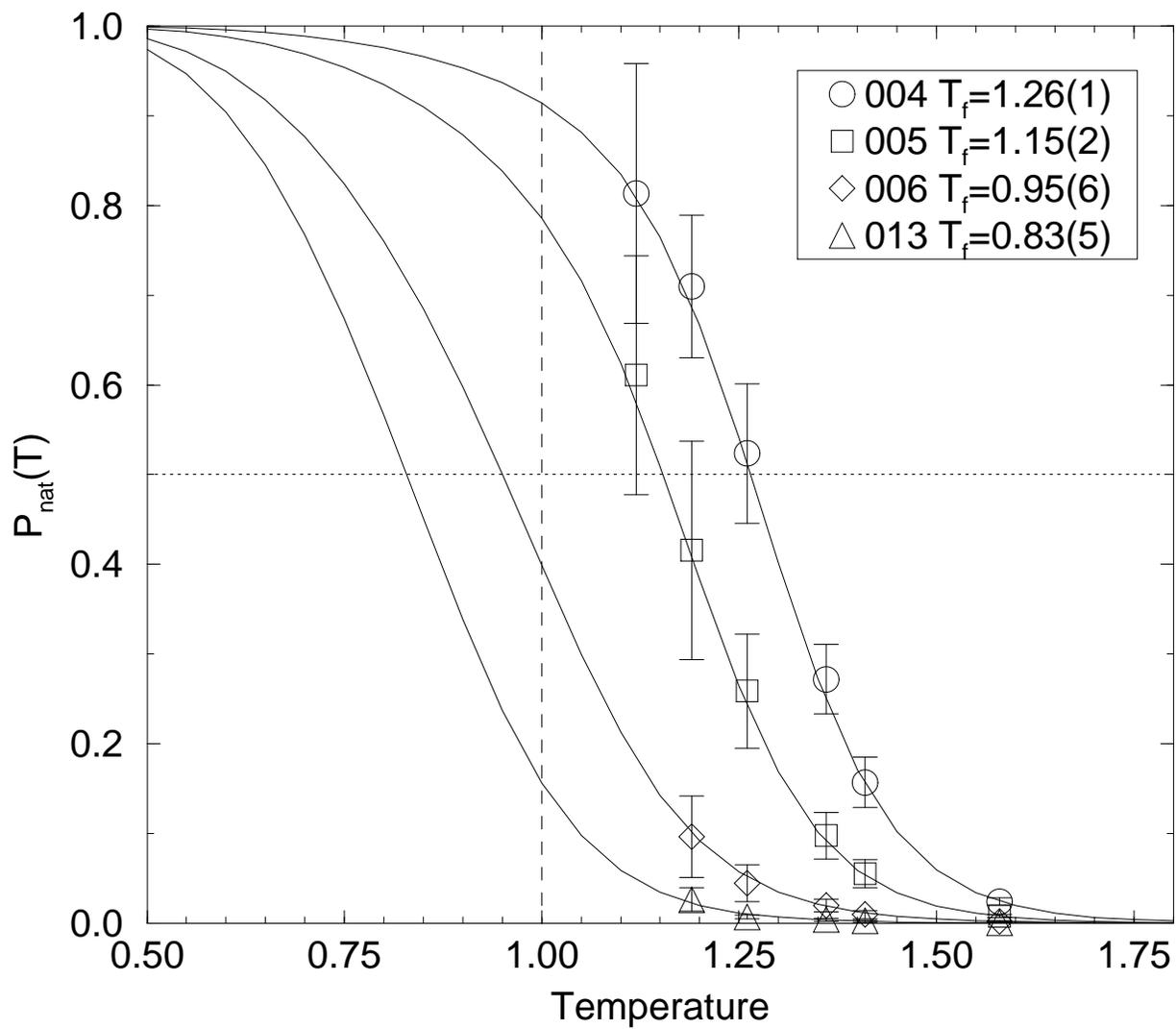
\begin{figure}
\caption{$P_{\rm nat}(T)$ for several sequences. $T_f$
  is defined to be the temperature at which $P_{\rm nat}(T_f)=.5$ The
  points were calculated using the standard Monte Carlo procedure.
  The solid lines are {\em not} fits to the data. They were calculated
  using the Monte Carlo Histogram Method,\protect\cite{Ferrenberg88}
  which enables one to calculate thermodynamic quantities at
  temperatures other than the simulation temperature. Two of the
  sequences have folding temperatures above $T_g$, the others have
  temperatures below $T_g$. The vertical line at $T=1$ indicates the
  glass transition point. The legend shows the folding temperatures,
  the numbers in the parenthesis indicate the uncertainty in the last
  digits.}
\label{fig:thermo}
\end{figure}

\end{document}